\documentclass[notitlepage,aps,prl,reprint,twocolumn,longbibliography,superscriptaddress]
{revtex4-1}
\usepackage{graphicx}
\usepackage{amsmath}
\usepackage{amssymb}
\usepackage{comment}
\usepackage[colorlinks, allcolors=blue]{hyperref}
\usepackage[all]{hypcap}
\usepackage[mathlines]{lineno}
\usepackage{physics}
\usepackage{wrapfig}
\usepackage{lipsum}
\usepackage[normalem]{ulem}
\usepackage{siunitx}
\usepackage{bbold}

\newcommand{\pref}[2]{\hyperref[#1]{\ref{#1}(#2)}}
\newcommand{\preff}[2]{\hyperref[#1]{\ref{#1}#2}}
\newcommand{\eqpref}[1]{\hyperref[#1]{(\ref{#1})}}

\newcommand{\squig}{{\raise.17ex\hbox{$\scriptstyle\sim$}}}

\begin{document}
	\title{Interaction-driven breakdown of Aharonov--Bohm caging in flat-band Rydberg lattices}
	\author{Tao Chen}
        \thanks{These authors contributed equally to this work.}
        \author{Chenxi Huang}
        \thanks{These authors contributed equally to this work.}
	\affiliation{Department of Physics, University of Illinois at Urbana-Champaign, Urbana, IL 61801-3080, USA}
	\author{Ivan Velkovsky}
	\affiliation{Department of Physics, University of Illinois at Urbana-Champaign, Urbana, IL 61801-3080, USA}
 \author{Tomoki Ozawa}
\email{tomoki.ozawa.d8@tohoku.ac.jp}
\affiliation{Advanced Institute for Materials Research (WPI-AIMR), Tohoku University, Sendai 980-8577, Japan}
 \author{Hannah Price}
\email{H.Price.2@bham.ac.uk}
\affiliation{School of Physics and Astronomy, University of Birmingham, Edgbaston, Birmingham B15 2TT, United Kingdom}
	\author{Jacob P. Covey}
 \email{jcovey@illinois.edu}
	\affiliation{Department of Physics, University of Illinois at Urbana-Champaign, Urbana, IL 61801-3080, USA}
	\author{Bryce Gadway}
\email{bgadway@psu.edu}
	\affiliation{Department of Physics, University of Illinois at Urbana-Champaign, Urbana, IL 61801-3080, USA}
 \affiliation{Department of Physics, The Pennsylvania State University, University Park, Pennsylvania 16802, USA}
	\date{\today}
 
\begin{abstract}
Flat bands play a central role in hosting emergent states of matter in many condensed matter systems,
from the nascent insulating states of twisted bilayer graphene to the fractionalized excitations found in frustrated magnets and quantum Hall materials.
Here, we report on the experimental realization of highly tunable flat-band models populated by strongly interacting Rydberg atoms. Using the approach of synthetic dimensions, we engineer a flat-band rhombic lattice with twisted boundaries, and through nonequilibrium dynamics we explore the control of Aharonov--Bohm (AB) caging via a tunable $U(1)$ gauge field.
Through microscopic measurements of Rydberg pairs, we explore the interaction-driven breakdown of AB caging in the limit of strong dipolar interactions that mix the lattice bands.
In the limit of weak interactions, where caging remains intact, we observe an effective magnetism that arises due to the interaction-driven mixing of degenerate flat-band states.
These observations of strongly correlated flat-band dynamics open the door to explorations of new emergent phenomena in synthetic quantum materials.
\end{abstract}

\maketitle

Frustration, resulting in degenerate eigenstates and flat energy bands that are sensitive to perturbations, underlies many of the phases and phenomena that define the forefront topics of condensed matter physics. This includes the fractionalized quasiparticles of quantum Hall matter~\cite{Stormer-FQHE} and spin models~\cite{Balents2010}, the intertwined orders of heavy fermion compounds~\cite{Coleman}, and the rapidly growing field of Moir{\'e} materials~\cite{MoireReview}.
Flat band lattices~\cite{Leykam-Flach-review}, where the frustration of electronic wavefunctions or interacting spins leads to perfectly degenerate energy bands, have played a specifically important role in enriching the understanding of itinerant ferromagnetism~\cite{Tasaki-3D-Ferro-flatband,Mielke_1991}
and lattice analogs~\cite{Sondhi-top-flat-bands,Torma-Superfluidity-flatband-top} of fractional Hall states~\cite{Stormer-FQHE,Jain-FQHE-theory}.

Recently, researchers have used the tools of synthetic quantum matter to engineer frustration~\cite{Leykam-Flach-review,Leykam-photonics} in photonic~\cite{Photonic-flat-band} and atomic~\cite{Taie-freezing,Takahashi-flat-bands,FlatBand-DSK-MF,Dawei-flat,li2022abcaging,Godfather-InverseAnderson} systems. The tunability of such platforms has even enabled the realization of Aharonov-Bohm (AB) caging~\cite{Thomson-caging-expt,AB-Christo,li2022abcaging,Vidal-caging-1,Longhi:14},
a condition where all bands become perfectly flat in the presence of a gauge field due to 
destructively interfering tunneling pathways.
Under AB caging, delocalized Bloch waves are transformed into compact localized states~(CLSs).

While many exciting questions relate to how interactions can lead to emergent physics in an AB caged flat-band lattice~\cite{Vidal-2-particle,Vidal-Caging-DisInt,Caging-Nonlinear,Flach-Caging-AllFlat}, realizations with light and atoms have been restricted to the non-interacting limit. Recently, experiments with superconducting qubits and Rydberg atoms have begun to probe interaction effects on a single frustrated plaquette~\cite{martinez2023flatband,chen2024}. Here, we explore the breakdown of AB caging in a flat-band lattice due to dipolar Rydberg interactions.
We engineer tunable flat band tight-binding models via microwave-driven Rydberg synthetic lattices~\cite{Kanungo2022,chen2024,Lu2024,Hazz-SynthDim-Rev}. For single atoms, we directly observe the phenomenon of AB caging and the independence of dynamics on the twist phase. For pairs, we observe the predicted~\cite{Vidal-2-particle} breakdown of AB caging due to interactions, with interaction-enabled delocalization for intermediate interaction strengths and the slow dynamics of bound pairs in the limit of strong interactions. Finally, we characterize the emergent magnetism of flat-band pseudospin states due to dipolar interactions.

\begin{figure*}[t!]
	\includegraphics[width=0.98\textwidth]{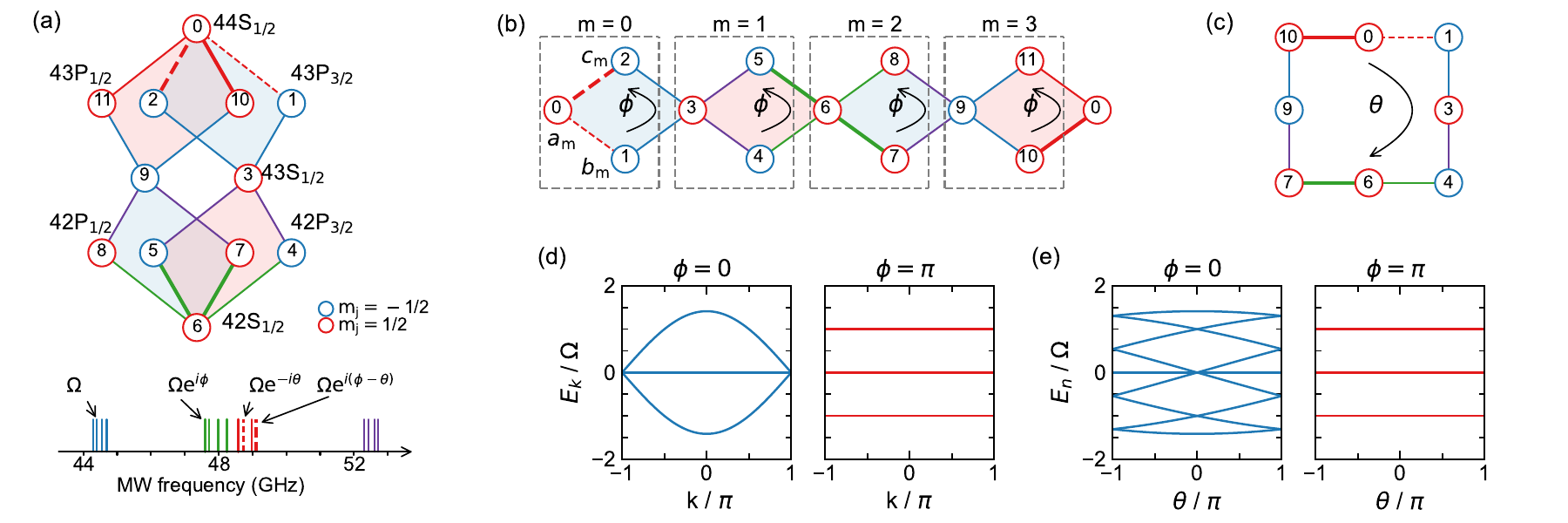}
	\caption{\textbf{Implementation of a twisted rhombic flat-band lattice in Rydberg synthetic dimensions.}
        \textbf{(a)}~A set of Rydberg states (top) are coupled with engineered multi-tone microwaves (bottom) to form a twisted rhombic structure. The red and blue circles correspond to $m_j=1/2$ and $-1/2$ sublevels in each $S$ or $P$ Rydberg state manifold, respectively. Both the plaquette flux $\phi$ and twist phase $\theta$ are introduced by tuning the phases of the microwave components driving the indicated (by bold and dashed lines) transitions relative to those driving the other transitions (thin solid lines). The coupling strength for each transition is calibrated by pairwise Rabi dynamics~\cite{SuppMats} and set to a common value $\sim\Omega/2$.
        \textbf{(b,c)}~Expansion and side view of the twisted structure. The plaquette flux is controlled by adding an additional phase $\phi$ to four transitions ($|2\rangle\leftrightarrow|0\rangle$, $|10\rangle\leftrightarrow|0\rangle$, $|6\rangle\leftrightarrow|5\rangle$ and $|6\rangle\leftrightarrow|7\rangle$, bold lines) relative to that of the other transitions in each plaquette. For $\phi=0$ and $\pi$, a twist phase $\theta$ can be tuned by introducing an additional phase $-\theta$ to the transitions $|1\rangle\leftrightarrow|0\rangle$ and $|2\rangle\leftrightarrow|0\rangle$.
        \textbf{(d)}~Eigenenergy bands of the extended rhombic lattice for plaquette flux $\phi=0$ (left) and $\pi$ (right) vs. quasimomentum $k$. 
        \textbf{(e)}~Fourfold-folded eigenenergy spectrum of the twisted rhombic lattice for $\phi=0$ (left) and $\pi$ (right) vs. the twist phase $\theta$.
}
\label{FIG:fig1}
\end{figure*}

We construct flat band lattices using the nascent approach of Rydberg synthetic dimensions~\cite{Kanungo2022,chen2024,Lu2024,Hazz-SynthDim-Rev,celi2014synthetic,Ozawa2019,Yuan:18}. Here, Rydberg states play the role of lattice sites, and the elements of an effective tight-binding model - the potential energy landscape and (complex) hopping terms - can be finely tuned through spectroscopic control over the transitions between the Rydberg levels~\cite{Hazz-SynthDim-Rev}. 
Dipolar interactions~\cite{Browaeys_2016} between neighboring Rydberg atoms in an array further introduces correlated pair-tunneling terms along the synthetic dimension~\cite{Sundar2018,chen2024}.

Extending earlier work~\cite{Kanungo2022,chen2024,Lu2024}, we utilize up to 12 Rydberg levels to engineer intricate 3D-like lattice structures with kinetic frustration and twisted~\cite{TBCs-Oded}, periodic boundary conditions as depicted in Fig.~\pref{FIG:fig1}{a}.
In this synthetic lattice, we explore both the single-atom and correlated-pair dynamics by preparing arrays of isolated atoms and isolated atom pairs. We prepare our Rydberg atom samples using the methods detailed in Refs.~\cite{JacksonPRR,chen2024,SuppMats}, based on the loading, cooling, and imaging of $^{39}$K atoms in optical tweezer arrays, followed by their excitation to the Rydberg level $nL_{J,m_J} = 42S_{1/2,1/2}$ [labeled by the synthetic ``site'' index 6 in Fig.~\pref{FIG:fig1}{a}].

Figure~\pref{FIG:fig1}{a} depicts how we create our twisted-boundary rhombic lattice structure by simultaneously driving 16 different microwave transitions between the magnetic sublevels ($m_J = \pm 1/2$ states) of the $S_{1/2}$, $P_{1/2}$, and $P_{3/2}$ manifolds for principal quantum numbers $n \in \{ 42,44\}$~\cite{SuppMats}. 
To make connection to the flat-band lattice model of interest, in  Fig.~\pref{FIG:fig1}{a} we label the twelve atomic levels that are used, and in Fig.~\pref{FIG:fig1}{b,c} we depict how these states relate to the unit cells and sub-lattice sites of the rhombic, or diamond, lattice~\cite{Vidal-2-particle,Thomson-caging-expt,Flach-perturbed,li2022abcaging}. This well-studied lattice possesses a unit cell with three sublattice sites, which we label $a$, $b$, and $c$. Our twelve-state implementation of periodic boundary conditions thus contains four unit cells, which we label by an index $m \in [0,3]$.
By controlling the relative phases of the applied microwave tones, we control the values of a uniform $U(1)$ Abelian flux $\phi$ that penetrates each rhombic plaquette as well as the ``twist phase'' $\theta$ associated with twisted boundary conditions (TBCs)~\cite{TBCs-Oded}, shown in Fig.~\pref{FIG:fig1}{b,c}.

\begin{figure}[t!]
	\includegraphics[width=0.5\textwidth]{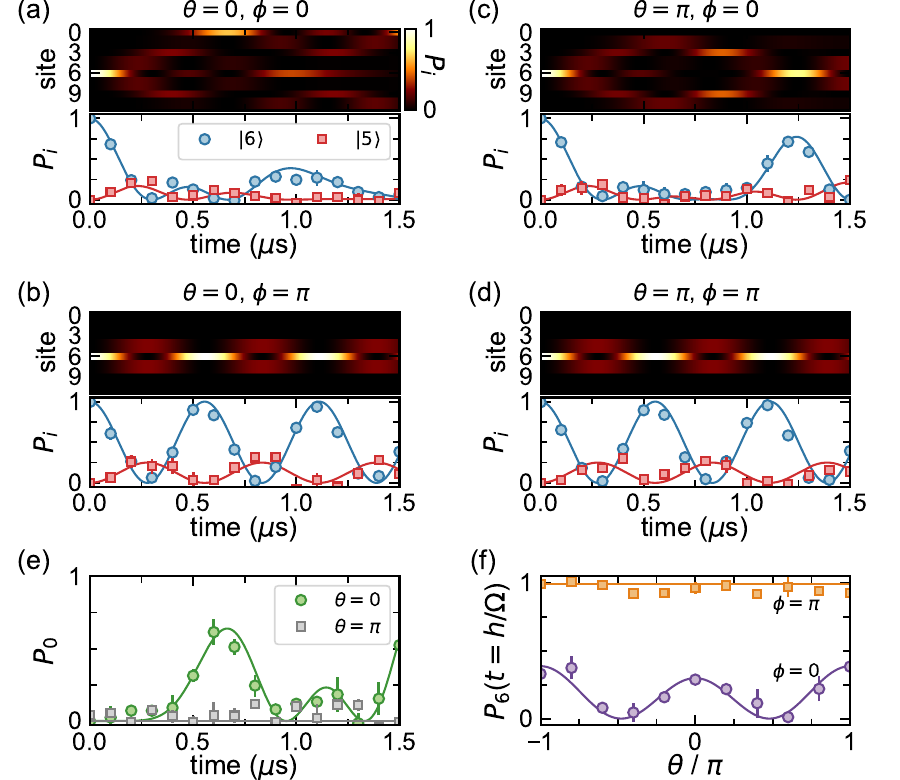}
	\caption{\textbf{Phase-dependent single atom caging dynamics on a synthetic rhombic lattice.}
        \textbf{(a-d)}~Top panels: Time evolution of the site (Rydberg state) populations
        from numerical simulations of
        Eq.~\ref{diamond-Ham}
        under different combinations of the twist phase $\theta$ and plaquette phase $\phi$.
        Bottom panels: Experimentally measured population dynamics (corrected for state preparation and measurement [SPAM] errors~\cite{SuppMats}) for states $|5\rangle$ (red squares) and $|6\rangle$ (blue circles) under the same phase combinations. Here the coupling strength is $\Omega/h=0.90(2)~{\rm MHz}$.
        \textbf{(e)}~Time evolution of the population in state $|0\rangle$ for $\theta=0$ (green circles) and $\pi$ (gray squares), for plaquette phase $\phi=0$.
        \textbf{(f)}~Population at the initial state $|6\rangle$ after dynamics for a time
        $t=h/\Omega$~$\sim$~1.1~$\mu{\rm s}$ as a function of the twist phase $\theta$, for $\phi=0$ (purple circles) and $\pi$ (orange squares), respectively. 
        Solid lines are numerical simulations with the ideal tight binding Hamiltonian. Error bars are the standard error from multiple independent sets. 
}
\label{FIG:fig2}
\end{figure}

We implement the rhombic lattice Hamiltonian
\begin{eqnarray}
    \hat{H} &=& \frac{\Omega}{2}\sum\limits_{m\in[0,3]} \left[ e^{i\phi^{(1)}_{ab,m}} \hat{b}_m^\dagger \hat{a}_m + e^{i\phi^{(1)}_{ca,m}} \hat{a}_m^\dagger \hat{c}_m \right. \nonumber\\ 
    ~ &+& \left. e^{i\phi^{(2)}_{ba,m}} \hat{a}^\dagger_{m+1} \hat{b}_m  + e^{i\phi^{(2)}_{ac,m}} \hat{c}_m^\dagger \hat{a}_{m+1} \right] + {\rm h.c.} 
\label{diamond-Ham}
\end{eqnarray}
with $\hat{a}_4 = \hat{a}_0$ for TBCs. The in-diamond plaquette phase $\phi$ and the twist phase $\theta$ are set by letting $\phi^{(1)}_{ab,0} = \theta$, $\phi^{(1)}_{ca,0} = \phi - \theta$, $\phi^{(1)}_{ab,2} = \phi$, $\phi^{(2)}_{ac,1} = \phi$ and $\phi^{(2)}_{ba,3} = \phi$, while the other phases are set to zero.
The tunneling energies are set to a uniform 
value of $\Omega/2$ 
by calibration and control of the MHz-scale state-to-state Rabi rates~\cite{SuppMats}. The uniform but tunable flux values $\phi$ are calibrated by measurement of the single-atom dynamics for isolated plaquettes. Finally, the twist angle $\theta$ is controlled by the intracell tunneling phases of the first unit cell, as indicated by the dashed lines in Fig.~\pref{FIG:fig1}{a-c}. TBCs~\cite{TBCs-Oded} allow for the exploration of effectively large systems with translation invariance, by using the twist phase $\theta$ to encode the phase accumulated by Bloch states with quasimomentum $k$.
This correspondence is depicted in Figs.~\pref{FIG:fig1}{d,e}, revealing the dispersion of eigenenergies with $k$ and $\theta$ when $\phi = 0$, contrasted with the flat response of all bands due to AB caging when $\phi = \pi$. 
The (e) energy dispersion of the (4 unit cell) TBC lattice vs. $\theta$ is just the fourfold-folded version of the (d) infinite lattice dispersion vs. $k$.

Experimental single-atom dynamics in the twisted rhombic lattice are presented in Fig.~\ref{FIG:fig2}. Starting with atoms prepared in state 6, the microwave-driven dynamics reveal our flux-based control of single-atom AB caging. In Fig.~\pref{FIG:fig2}{a-d}, we show the full set of simulated state dynamics, along with the measured and simulated dynamics for the initialized state ($\ket{6}$, blue circles) and a neighboring state ($\ket{5}$, red squares). Extended data with the dynamics of more states are presented in~\cite{SuppMats}.

The dynamics under non-caged ($\phi = 0$) and caged ($\phi = \pi$) conditions are first contrasted in Fig.~\pref{FIG:fig2}{a,b} for zero twist phase.
In the absence of caging ($\phi = 0$), population spreads out from the initial unit cell. In contrast, under AB caging ($\phi = \pi$), population oscillates between state 6 and the other states of the central unit cell.
Here, the initial population projects onto just 3 nearby CLSs that beat against each other leading to intracell dynamics but no large-scale delocalization.
These two trends are reflected in the measured population dynamics for states $\ket{5}$ and $\ket{6}$ in Fig.~\pref{FIG:fig2}{a,b}, which are in good agreement with the simulations and show the emergence of large, regular population revivals under caging.

Figure~\pref{FIG:fig2}{c,d} present corresponding simulations and state population dynamics, but for a twist phase of $\theta = \pi$. While the dynamics under caging in Fig.~\pref{FIG:fig2}{d} are identical to those in Fig.~\pref{FIG:fig2}{b}, the change of twist phase from 0 to $\pi$ leads to significant modifications to the non-caged dynamics in Fig.~\pref{FIG:fig2}{c}.
Microscopically, a $\pi$ twist phase around the periodic boundary structure of Fig.~\pref{FIG:fig1}{c} can be understood to cause an additional, larger-scale caging condition, wherein population starting at site $\ket{6}$ is forbidden from reaching the opposing site, $\ket{0}$, due to destructive interference.
This results in the enhanced revival at site $\ket{6}$ after a time
$\sim$1.2~$\mu$s 
seen in Fig.~\pref{FIG:fig2}{c}.

We probe this larger-scale caging condition more directly in Fig.~\pref{FIG:fig2}{e} by measuring the site $\ket{0}$ population dynamics under a twist phase of $0$ (green circles) and $\pi$ (gray squares). Indeed, following initialization at site $\ket{6}$, we find that population reaches to $\ket{0}$ in good agreement with the theory under zero twist phase, while very little population is seen to reach $\ket{0}$ for $\theta = \pi$.
In Fig.~\pref{FIG:fig2}{f} we probe another key feature of the twist phase $\theta$, namely the independence of the nonequilibrium dynamics on $\theta$ under AB caging conditions.
We specifically measure the population remaining at the initialized site $\ket{6}$ after a time of $1.1~\mu$s,
corresponding to two oscillations in the
case of caging
($t \sim h/\Omega$).
We find that the dynamics are dependent on $\theta$ for $\phi = 0$ (purple circles), reflecting the dispersive energy bands under non-caging conditions in Fig.~\pref{FIG:fig1}{d,e}, whereas the measurements are essentially independent of $\theta$ for the caging condition ($\phi = \pi$, orange squares), consistent with the all-flat-bands condition.

\begin{figure*}[t!]
	\includegraphics[width=0.98\textwidth]{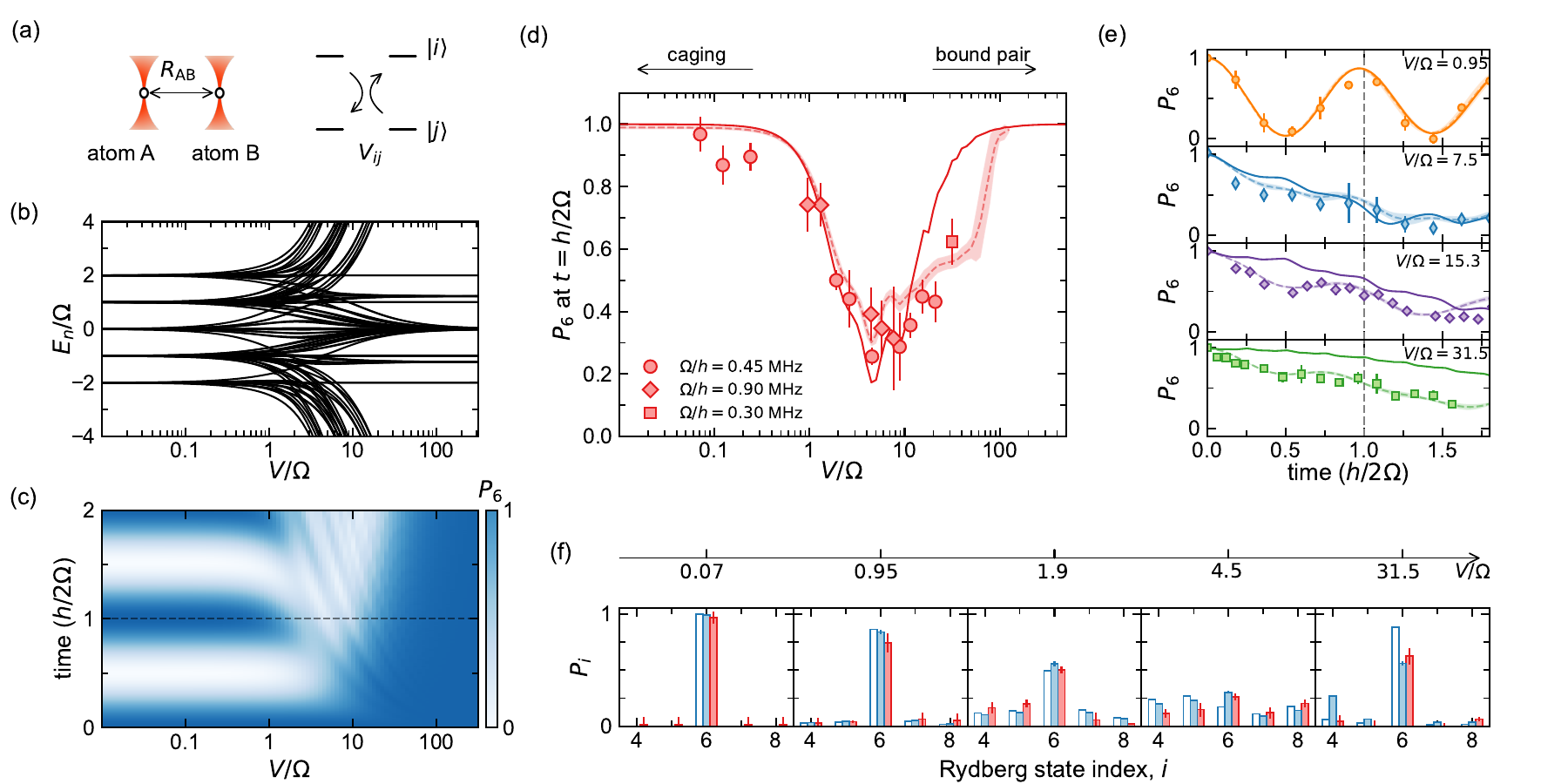}
	\caption{\textbf{Breakdown of AB caging by interactions and the crossover to bound pair dynamics.}
        \textbf{(a)}~Atom pair $A$ and $B$ prepared in tweezers with a spatial separation of $R_{\rm AB}$. The dipolar exchange interaction for state $\ket{i}$ and $\ket{j}$ is $V_{ij}$, and we scale all interactions to $V=V_{67}$ with calculated $C_3$ coefficients~\cite{SuppMats}.
        \textbf{(b)}~Eigenenergy distribution for atom pairs in the 12-state rhombic lattice (with $\phi=\pi$ and $\theta=0$) under different interaction-to-coupling ratios $V/\Omega$.
        \textbf{(c)}~Time evolution of the population in $\ket{6}$, $P_6$, for different $V/\Omega$ ratios from ideal numerical simulations. The dashed line indicates the single-atom revival time. 
        \textbf{(d)}~Measured crossovers from caging, to delocalization, to pairs with reduced mobility.
        \textbf{(e)}~Experimentally measured time evolution of the SPAM-corrected population in $\ket{6}$ for different $V/\Omega$. The dashed line again corresponds to the single-atom revival time.
        \textbf{(f)}~Population distributions for the central 5 sites at the first revival time $t=h/2\Omega$ for different $V/\Omega$. The red and white bars correspond to experimental measurements and ideal numerical simulations, respectively. 
        The blue bars indicate the full numerical simulations that include all interactions
        (including state-changing terms), SPAM errors, and finite-temperature effects.
        Solid lines in (d,e) are simulations based on Eq.~\ref{diamond-Ham} with
        resonant flip-flop interactions, 
        while the dashed lines and their shaded confidence intervals are Monte Carlo simulations that also include state-changing interaction terms, thermal variations of $R_{\rm AB}$, along with SPAM errors.
        All error bars are the standard error of multiple independent data sets.
}
\label{FIG:fig3}
\end{figure*}

More generally, the observed AB caging and flat bands result from kinetic frustration. Generically, perturbations strong enough to mix the bands can disrupt frustration and induce delocalization~\cite{Vidal-Caging-DisInt}.
Recently, the onset of transport under the addition of strong disorder has been observed in flat-band lattices~\cite{li2022abcaging,Godfather-InverseAnderson}, relating to the phenomenon of inverse Anderson localization. More intriguingly, it has been predicted that interparticle interactions alone can lead to the emergent breakdown of flat-band localization~\cite{Vidal-2-particle}. While nonlinear modifications to flat lattice bands have recently been measured in bosonic quantum gases~\cite{Takahashi-flat-bands,FlatBand-DSK-MF}, the predicted delocalization of particles due to strong interactions is still outstanding.

\begin{figure*}[t!]
	\includegraphics[width=0.95\textwidth]{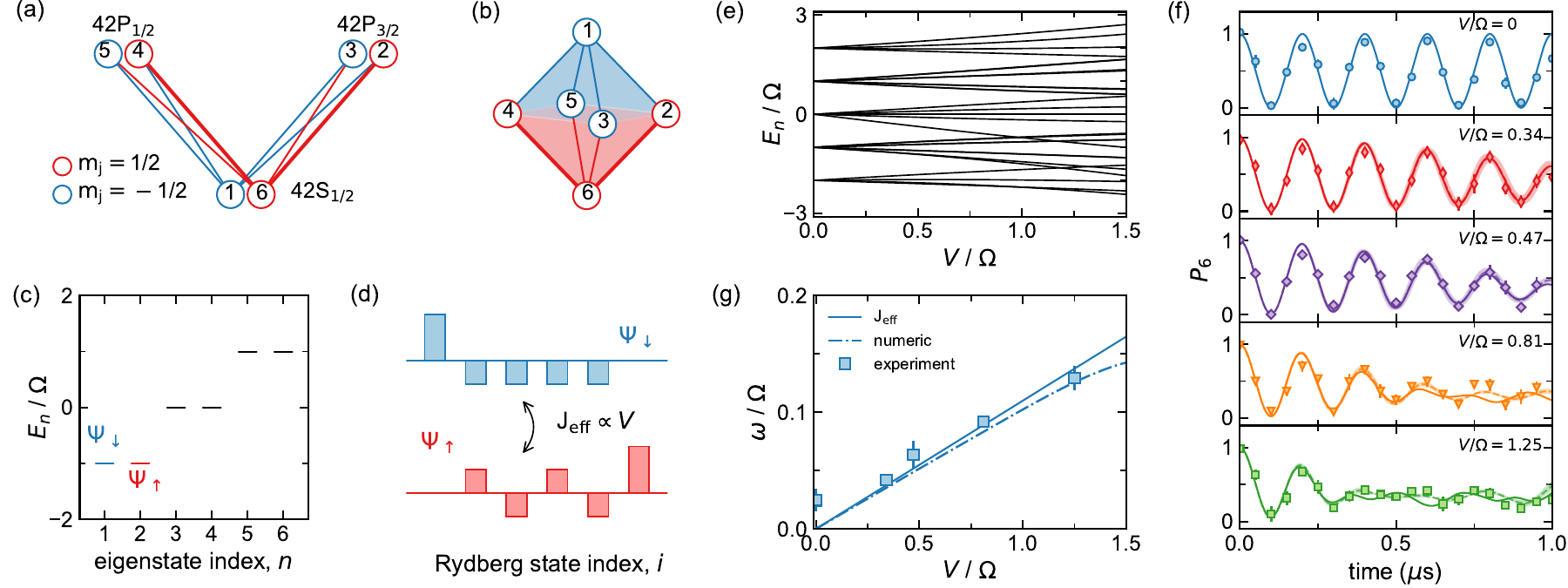}
        \centering
	\caption{\label{FIG:fig4}
        \textbf{Emergent magnetism
        in a flat-band $\pi$-flux rhombic bipyramid.}
        \textbf{(a)}~Six Rydberg levels are coupled by 8 microwave tones to generate the \textbf{(b)} bipyramid structure. The two connections for $\ket{6}\leftrightarrow\ket{2}$ and $\ket{6}\leftrightarrow\ket{4}$, indicated by bold lines, have a $\pi$ phase shift relative to the other transitions.
        \textbf{(c)}~Doubly-degenerate eigenspectrum of the non-interacting 6-state system. We denote the degenerate pairs of eigenstates by effective pseudospins $\Psi_\downarrow$ (blue) and $\Psi_\uparrow$ (red), shown for the lowest two states.
        \textbf{(d)}~Representation of the lowest degenerate single-atom eigenstates $\Psi_\uparrow$ (red) and $\Psi_\downarrow$ (blue). In the presence of weak dipolar exchange ($V \ll \Omega$), the emergent magnetic interactions $J_{\rm eff}$ are proportional to $V$ (scaled to $V=V_{62}$)~\cite{SuppMats}.
        \textbf{(e)}~Atom pair eigenenergies for increasing $V/\Omega$.
        \textbf{(f)}~Time evolution of the SPAM-corrected mean population in $\ket{6}$ for different $V/\Omega$ ratios. The coupling strength is $\Omega/h=2.50(4)~{\rm MHz}$, while the interaction strength $V$ is varied via the atomic separation.
        The solid lines are numerical simulations with experimental parameters.
        The dashed lines and shaded regions are numerical simulations that consider both the finite-temperature spread of interactions and the Rydberg state preparation infidelity.
        \textbf{(g)}~Short time beating frequency $\omega$ versus the interaction strength, found by fitting the experimental measurements over the first 0.5~$\mu{\rm s}$ to the function $P_6(t)=a + b\cos{(2\omega t/\hbar)}\cos{(2\Omega t/\hbar)}$. The solid line is the analytical prediction, $J_{\rm eff}$, based on the emergent magnetism in the $\{ \Psi_\downarrow, \Psi_\uparrow \}$ subspace~\cite{SuppMats}.
        The dotted dash line is from fitting to the numerical results in (f), over the same time with the same function.
        All error bars are standard errors from multiple experimental measurements.
}
\end{figure*}

Using pairs of Rydberg atoms with dipole-dipole interactions, we now explore the dynamics of strongly interacting pairs in the same flat-band lattice from Figs.~\ref{FIG:fig1} and \ref{FIG:fig2}.
As depicted in Fig.~\pref{FIG:fig3}{a}, we study pairs of atoms, labeled $A$ and $B$, spaced by a tunable distance $R_{\rm AB} > 4$~$\mu$m, relating to tunable MHz-scale interactions.
The atoms primarily interact through resonant dipolar exchange~\cite{Browaeys_2016}, which relates to anti-correlated hopping of the $A$ and $B$ Rydberg electrons along the synthetic lattice~\cite{Sundar2018,chen2024}, characterized by a rate $V_{ij} \propto C^{ij}_3/2R^3_{\rm AB}$ for the transition $\ket{i}_{\rm A} \ket{j}_{\rm B} \leftrightarrow \ket{j}_{\rm A} \ket{i}_{\rm B}$ (with $C^{ij}_3$ the state-dependent $C_3$ coefficient)~\cite{SuppMats}. These interactions lack translation symmetry along the synthetic dimension due to the dependence of the exchange rates on the participating Rydberg levels~\cite{SuppMats}. For simplicity, we characterize the $A$-$B$ interactions by a single scale $V$ relating to the rate $V_\textrm{67}$ of $\ket{6}_{\rm A}\ket{7}_{\rm B} \leftrightarrow \ket{7}_{\rm A}\ket{6}_{\rm B}$ exchange.

For moderate strengths, despite their structure along the synthetic dimension, we expect that the interactions will generically disrupt AB caging and induce transport. This is reflected in the energy spectrum in Fig.~\pref{FIG:fig3}{b}, where for $V/\Omega \sim 1$ the flat, isolated energy bands become strongly mixed. The resulting reorganization of the system eigenstates will also be reflected in the nonequilibrium dynamics. Figure~\pref{FIG:fig3}{c} shows the expected dynamics of the mean state $\ket{6}$ population, $P_6$, as a function of $V/\Omega$, for an initial product state $\ket{6}_{\rm A}\ket{6}_{\rm B}$. Similar to the energy spectrum, three regimes are observed: robust revivals due to AB caging for $V/\Omega \lesssim 1$, a decay due to delocalization dynamics for $1 \lesssim V/\Omega \lesssim 20$, and a freeze-out of $P_6$ dynamics for very large $V/\Omega$ due to
the inhibition of hopping
by strong, nearly random interactions~\cite{SuppMats,Sierant}.

In Fig.~\pref{FIG:fig3}{d} we experimentally measure $P_6$ for interacting atom pairs at the first single-atom revival time ($t = \hbar / 2\Omega$, dashed line in Fig.~\pref{FIG:fig3}{c}), using several values of the spacing $R_{\rm AB}$ and the tunneling strength to vary $V/\Omega$ by more than two orders of magnitude. We observe the expected disruption of AB caging for intermediate interactions, in good agreement with the ideal simulations (solid line) based on Eq.~\ref{diamond-Ham} and resonant dipolar exchange interactions. For strong interactions, we find that  the measured $P_6$ rises back up, in qualitative agreement with the expectations based on Figs.~\pref{FIG:fig3}{b,c}. For the largest interactions ($V/\Omega \gtrsim 10$), the data begins to lose agreement with the idealized description of resonant exchange interactions that conserve the net ($A$ and $B$ combined) populations of each Rydberg level. Better agreement is found if we account for the expected contributions from dipolar interaction terms that interconvert spin and orbital angular momentum~\cite{Browaeys_2016}, represented by the dashed curve in Fig.~\pref{FIG:fig3}{d}. With the state-changing dipolar interactions included, a full freeze-out of the $P_6$ dynamics is not expected until inaccessibly large interactions ($V/\Omega \gtrsim 100$).
The dashed lines also account for minor sources of parameter uncertainty, i.e., from calibration uncertainty and the thermal spread of interactions~\cite{SuppMats}.

In Fig.~\pref{FIG:fig3}{e}, we plot traces of the measured dynamics for a few representative values of $V/\Omega$,
finding good agreement with the expected pair dynamics (dashed curves).
We can identify two main sources of disagreement between the data and the simple idealized model (solid lines). For intermediate $V/\Omega$ ($V/\Omega = 7.5$ panel), the disagreement arises mainly from the non-negligible contribution of single atoms. With a 92$\%$ Rydberg state preparation fidelity (combination of optical pumping and STIRAP efficiencies), roughly 15$\%$ of our ``pair'' data contains only a single Rydberg atom. This is accounted for in the dashed curve by weighting the expected dynamics of singles and pairs. For large interactions, $V/\Omega \gtrsim 10$, the non-resonant state-changing interactions become the dominant source of disagreement.

In Fig.~\pref{FIG:fig3}{f}, we separately read out the population from several different Rydberg states to further probe the predicted interaction-induced breakdown of AB caging. After an evolution for $t$~$=$~$\hbar/2\Omega$, we read out the population at different Rydberg levels by performing state-swapping microwave transitions prior to optical de-excitation of $42S_{1/2,1/2}$. We plot histograms of the populations for states $\ket{4}$-$\ket{8}$ (the initialized site and its neighbors), comparing the measured populations (red bars) to those predicted by the full (blue) and ideal (white) simulations. In general, we find good evidence for significant population residing on site $\ket{6}$ for both small and large $V/\Omega$, with population spread across the set of measured states for intermediate $V/\Omega$ values.
While we focus on the delocalization of just two interacting particles, Rydberg synthetic dimensions can be extended to hundreds of atoms~\cite{Ebadi2021}
and many dozens of states, providing a versatile complement to explorations of correlated quantum walks with neutral atoms~\cite{Preiss-QWalk,Kaufman-QWalk} and photons~\cite{OBrien-QWalk}.

Our observation of the interaction-induced breakdown of AB caging in Fig.~\ref{FIG:fig3} occurred in the $V > \Omega$ regime, when pairwise interactions are so large that they mix the bands and destroy flat band localization.
In contrast, when $V < \Omega$, weak interactions do not directly mix the bands, but they act as a first-order perturbation to the physics of the individual, isolated bands. This regime results in rich, emergent physics driven by interactions. In generic flat bands occupied by short-range-interacting particles, one encounters emergent long-range interactions and phenomena like charge-density-wave ordering~\cite{HuberAltman}. More exotic phenomena are encountered in topological flat bands~\cite{Torma-FlatBand-Superfluidity}, including analogs of the fractional quantum Hall effect in lattice systems~\cite{Sondhi-top-flat-bands}. In Fig.~\ref{FIG:fig4}, we use a simple flat band structure to explore the emergent magnetism of flat band localized states.

Figure~\pref{FIG:fig4}{a,b} depict the 8 Rydberg levels we utilize and the geometric representation of the engineered flat band structure. Solid lines denote the hopping links, and a $\pi$ hopping phase is applied on the two thick lines ($\ket{4} \leftrightarrow \ket{6}$ and $\ket{6} \leftrightarrow \ket{2}$). The resulting structure is a three-dimensional rhombic bipyramid that has $\pi$ flux penetrating each of its four surfaces, which can be thought to result from an enclosed magnetic monopole. This leads to an Aharonov--Bohm caging, with three degenerate pairs of CLSs. This is exemplified in Fig.~\pref{FIG:fig4}{c,d}, where we label the ground pair of degenerate states as $\Psi_\downarrow$ and $\Psi_\uparrow$. The result of weak interactions between neighboring Rydberg atoms (referred to again as $A$ and $B$) is an emergent spin-$1/2$ quantum magnetism of these degenerate pseudospin state pairs~\cite{SuppMats}, reflected in the linear splitting of 2-atom eigenstates in Fig.~\pref{FIG:fig4}{e} with increasing interactions.

In the idealized picture of uniform interactions between nearest synthetic neighbors ($V_{6j}=V_{1j}=V$ for $j\in [2,5]$), the low-energy physics is effectively described by the spin model $J_{\rm XX}(\sigma_{\rm A}^x\sigma_{\rm B}^x + \sigma_{\rm A}^y\sigma_{\rm B}^y) + J_{\rm ZZ}(\sigma_{\rm A}^z\sigma_{\rm B}^z)$, where $J_{\rm ZZ} = J_{\rm XX} = V/4$, relating to the well-studied Heisenberg~XXX model.
Considering the state-specific form of the $V_{ij}$ in our system, the actual description includes an additional spin-spin interaction term, $J_{\rm XZ}(\sigma_{\rm A}^x\sigma_{\rm B}^z + \sigma_{\rm A}^z\sigma_{\rm B}^x)$. With the experimental interaction terms $\{V_{62}, V_{63}, V_{64}, V_{65}\} = \{V, -V/4, V/2, -V/2\}$ and $\{V_{12}, V_{13}, V_{14}, V_{15}\} = \{-V/4, V, -V/2, V/2\}$ (with $V \equiv V_{62}$), we have $J_{\rm XZ} = 3J_{\rm ZZ} = 3J_{\rm XX} = 9V/64$. To probe this emergent magnetism, we simply monitor the dynamical response of atom pairs initialized to the state $\ket{6}_{\rm A}\ket{6}_{\rm B}$. Figure~\pref{FIG:fig4}{f} shows the dynamics of the mean $P_6$ for several interaction-to-hopping ratios $V/\Omega$. For weak $V/\Omega$, we observe two well-separated contributions to the dynamics: a fast oscillation and a slower beating that leads to an apparent contrast decay. The fast term, $\sim$$\Omega$, relates to the intra-cell dynamics that stem from the beating of the different energy bands. The low-frequency part results from the interaction-induced splitting of the individual bands, and relates directly to the emergent magnetism. For our physical interactions, we expect it to scale approximately as $J_{\rm eff}=\sqrt{J_{\rm XX}^2 + J_{\rm XZ}^2/2}$.
Figure~\pref{FIG:fig4}{g} shows the fit-determined low-frequency contribution, $\omega$. This emergent oscillation term scales roughly linearly with $V$ for small $V/\Omega$, in good agreement with the $J_{\rm eff}$ prediction and with the prediction based on numerical simulations, which begins to deviate from a linear ($\propto V$) relationship at large $V$ due to band mixing.

These explorations of strongly interacting particles in flat-band lattices
suggest several future extensions.
Using the many states available to Rydberg synthetic lattices, it is possible to explore the related phenomena of non-Abelian AB caging~\cite{NA-AB-Caging} as well as AB caging in higher (3 or even 4) dimensions.
Extending to topological flat band models, we may ask whether the non-local (in space) interactions can result in emergent topological order.
Finally, in the context of pseudo-magnetism emerging from the projection of interactions onto flat-band CLSs, higher-spin models will emerge from lattices with more unit cells (e.g., a spin-3/2 model emerges from the four-cell rhombic lattice), providing a route to explore, e.g., emergent
quantum Potts models~\cite{PottsRG,Cohen-clock}.

\section{Acknowledgements}
We thank Tabor Electronics greatly for the use of an arbitrary waveform generator demo unit. This material is based upon work supported by the National Science Foundation under grant No.~1945031 and the AFOSR MURI program under agreement number FA9550-22-1-0339. TO is supported by JSPS KAKENHI Grant No. JP20H01845, JST PRESTO Grant No. JPMJPR2353, JST CREST Grant No. JPMJCR19T1. HMP is supported by the Royal Society via grants UF160112, URF\textbackslash R\textbackslash221004, RGF\textbackslash E A\textbackslash180121 and RGF\textbackslash R1\textbackslash180071, and by the Engineering and Physical Sciences Research Council (grant no. EP/W016141/1). This work was also supported by the BRIDGE Seed Fund for collaboration between the University of Birmingham and the University of Illinois at Urbana-Champaign.

\bibliographystyle{apsrev4-1}
\bibliography{Ryd}

\clearpage

\renewcommand{\thesection}{\Alph{section}}
\renewcommand{\thefigure}{S\arabic{figure}}
\renewcommand{\thetable}{S\Roman{table}}
\setcounter{figure}{0}
\renewcommand{\theequation}{S\arabic{equation}}
\renewcommand{\thepage}{S\arabic{page}}
\setcounter{equation}{0}
\setcounter{page}{1}

\newpage

\begin{widetext}
\appendix

\section{\large Supplemental Material for ``Interaction-driven \\ breakdown of Aharonov--Bohm caging in flat-band Rydberg lattices''}

\vspace{5mm}

\section{ADDITIONAL EXPERIMENTAL DETAILS}

\subsection{Normalization of Rydberg state population dynamics -- corrections for preparation and readout infidelity}

As discussed in Ref.~\cite{chen2024}, the primary data we measure for the state populations appear similar to those presented in Fig.~\ref{FIG:figs1}. There are two limiting quantities to note - upper and lower limits to the data. First, there is an upper ceiling value that is on average equal to $P_u = 0.88(1)$, which stems from inefficiency of STIRAP and loss during release-and-recapture. There is also a lower baseline of the measurements, having an average value $P_l = 0.21(1)$, that we believe stems from the decay (and subsequent recapture) of the short-lived Rydberg states (here, $n \in \{42, 43, 44\}$), such that the Rydberg states have some probability to appear bright to subsequent fluorescence detection.
These infidelities limit the contrast of state population dynamics.

For all of the data in the paper, we renormalize to account for these known infidelities in the following way: we define the renormalized populations $P_i$ in relation to the measured bare populations $P_i^\textrm{bare}$ as $P_i = (P_i^\textrm{bare}-P_l)/(P_u-P_l)$. In the main text, we refer to such renormalized population measurements as being SPAM-corrected.

It is relevant to note that the statistical fluctuations associated with the processes that motivate this discussed renormalization are systematically not reflected in either the error bars of the renormalized data nor the error bars of the full theory, which only accounts for the calibrated uncertainties of the control parameters and the thermal fluctuations of the inter-atomic separations.

\subsection{Calibrations of synthetic lattice hopping amplitudes and plaquette fluxes}

\begin{figure*}[b]
	\includegraphics[width=\textwidth]{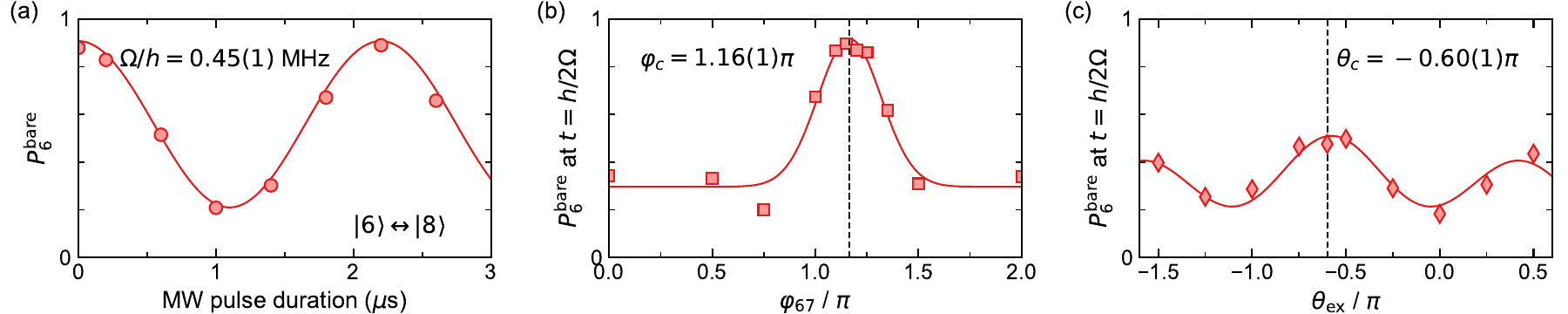}
        \centering
	\caption{\label{FIG:figs1}
        \textbf{Calibration of state-to-state coupling strength $\Omega$, in-diamond flux $\phi$, and twist phase $\theta$.}
        \textbf{(a)}~ Rabi oscillation for $\ket{6}\leftrightarrow\ket{8}$ transition. Fitting with a sine function (solid line) to the experimental data (circles) gives the coupling strength is $\Omega/h=0.45(1)~{\rm MHz}$.
        \textbf{(b)}~ Population in $\ket{6}$ after an evolution time of $t=h/2\Omega$ ($\sim 1.1~\mu$s) versus the relative source phase $\varphi_{67}$ for the $\ket{6}\to\ket{7}$ tone in the IF generator output signal. The Gaussian fit (solid line) gives the peak location $\varphi_c = 1.16(1)\pi$, which corresponds to zero in-diamond flux $\phi=0$.
        \textbf{(c)}~ Population in $\ket{6}$ after an evolution time of $t=h/\Omega$ ($\sim 1.1~\mu$s, $\Omega/h=0.90(2)~{\rm MHz}$) with in-diamond flux $\phi=0$ versus the additional phase $\theta_{\rm ex}$ for $\ket{1}\to\ket{0}$ and $\ket{2}\to\ket{0}$ transitions. By fitting the measured data with a 3-peak Gaussian function (peaks at $\theta_c$ and $\theta_c\pm \pi$), we get the calibrated twisted phase $\theta_c=-0.60(1)\pi$ that corresponds to zero twist phase.
}
\end{figure*}

Comparing to our previous demonstration of Rydberg synthetic dimensions in atom arrays~\cite{chen2024}, here we drive more transitions between more Rydberg levels. Most importantly, to address transitions between states with principal quantum numbers $n = 42$, 43, and 44, here we had to apply coherent microwaves over a greater span of frequencies. To address these transitions, here modify the setup used in Ref.~\cite{chen2024} to use a Tabor P9484D for the intermediate frequency (IF) signal, achieving a bandwidth of $\sim$9~GHz centered around a frequency of approximately 48~GHz. 

We calibrate the mapping between the parameters of our applied microwave spectrum and the those of the effective tight-binding model, Eq.~\ref{diamond-Ham}, based on the atomic response. The control of the parameters (tunneling amplitudes, plaquette fluxes, and twist phase) via the microwave tone parameters is discussed in the main text. In short, before expounding, we calibrate the hopping rates by simply measuring state-to-state Rabi dynamics. We calibrate the fluxes in each of the four primary lattice plaquettes by measuring the recurrence dynamics when beginning on one of the adjoining $A$ sublattice sites. And we similarly calibrate the twist phase by measuring the dynamical response.

Figure \pref{FIG:figs1}{a} shows an example calibration of state-to-state hopping rates. To measure the Rabi frequency between state $\ket{i}$ and $\ket{j}$, we first apply a series of high-fidelity $\pi$-pulses to transfer all population from initially prepared state $\ket{6}$ to state $\ket{i}$, then measure the $\ket{i}\leftrightarrow\ket{j}$ oscillation by varying the single-tone microwave pulse duration followed by another series of $\pi$-pulses (from $\ket{i}$ back to $\ket{6}$) and the detection ``pulse'' (975 nm laser, depumping from $\ket{6}$ to the ground $4S_{1/2}$ state for fluorescence imaging)~\cite{chen2024}. We vary the IF amplitudes for each specific frequency tone to make the Rabi frequencies for all relevant transitions nearly uniform (to a value we refer to as $\Omega/h$). Our microwave generation system works within the linear response regime, so we can simply scale the global output amplitude of the IF generator to globally change the value of $\Omega$ after calibration. Additionally, we note that the calibration of the individual links in the absence of other applied tones yields results that are consistent with the response in the presence of all tones (i.e., corrections due to off-resonant coupling terms from the other drives are small and ignorable).

\begin{figure*}[b]
	\includegraphics[width=\textwidth]{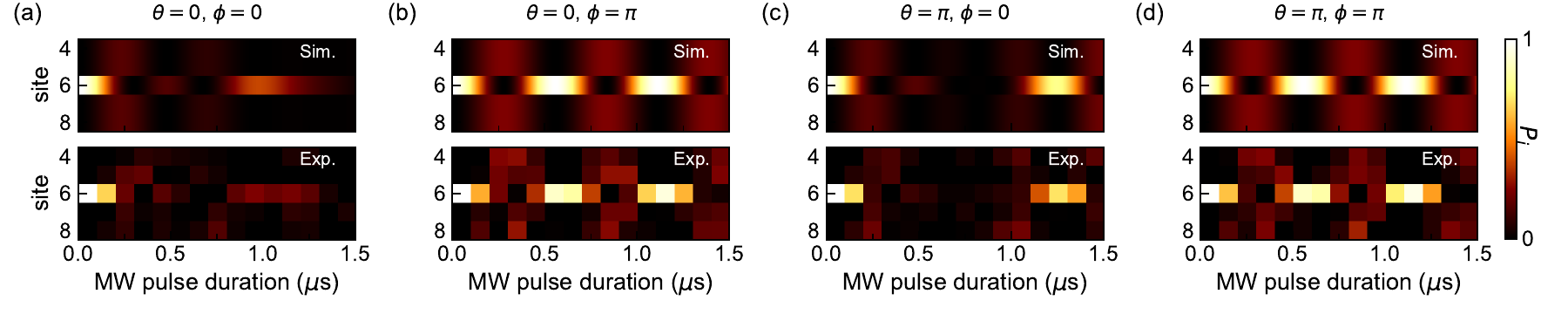}
        \centering
	\caption{\label{FIG:figs2}
        \textbf{Additional experimental data for the single-atom population dynamics of main-text Fig.~2, under different phase ($\theta$ and $\phi$) combinations, showing the simulated dynamics (top) and measured population dynamics (bottom) for the states 4-8.}
        \textbf{(a)}~Dynamics for $\phi=0$ and $\theta=0$.
        \textbf{(b)}~Dynamics for $\phi=\pi$ and $\theta=0$.
        \textbf{(c)}~Dynamics for $\phi=0$ and $\theta=\pi$.
        \textbf{(d)}~Dynamics for $\phi=\pi$ and $\theta=\pi$.
}
\end{figure*}

Figure \pref{FIG:figs1}{b} shows how we calibrate the in-diamond, or plaquette, flux $\phi$ for the $\ket{6}\to\ket{7}\to\ket{9}\to\ket{8}\to\ket{6}$ plaquette (see Fig.~1 of the main text for the state configuration mapping). We let the frequency tone (sine wave) in the IF generator output for the $\ket{i}\to\ket{j}$ transition take a phase of $\varphi_{ij}$. For this plaquette, we let $\varphi_{79} = \varphi_{98} = \varphi_{86} = 0$ and scan $\varphi_{67}$ from 0 to $2\pi$ to check the in-diamond phase-dependent dynamics (all these phases refer to the initial phase values at the source). Similar to Ref.~\cite{chen2024}, after an evolution time of $t=h/\Omega$, the population in $\ket{6}$ has a peaked recurrence for zero in-diamond phase. As shown in Fig.~\pref{FIG:figs1}{b}, the peak location at $\varphi_{67}=1.16(1)\pi$ indicates the in-diamond phase $\phi=0$ at the atoms. We then simply account for the the calibrated offset phase $\varphi_c=1.16\pi$ in the IF output program. We find that this calibrated offset phase value is very stable in our day-to-day measurements. For other plaquettes, we repeat a similar calibration procedure based on population recurrence (using a series of $\pi$ pulses to initiate population at, and read out population from, the states at the corner of those plaquettes).

After the calibrations of the state-to-state coupling strengths and each in-diamond flux, we address the twist phase $\theta$ by simply scanning the extra additional phases $-\theta_{\rm ex}$ on both the $\ket{1}\to\ket{0}$ and $\ket{2}\to\ket{0}$ transitions (see the main text - these phases cancel in their contributions to the $m = 0$ plaquette flux, but contribute to the global twist phase). We initiate population at state $\ket{6}$ and measure the remaining population, $P_6$, after the system evolves $t=\hbar / \Omega\sim 1.1$~$\mu$s) with $\Omega/h=0.9(2)~{\rm MHz}$. From the theoretical simulations, $P_6$ peaks at twist phase values of both 0 and $\pi$ at this time (when $\phi = 0$, under the non-caged condition, see solid curve Fig.~2(f) of the main text). As shown in Fig.~\pref{FIG:figs1}{c}, we fit the measured $P_6$ with a set of Gaussian peaks separated by $\pi$ phase and obtain that $\theta_{\rm ex} = \theta_c=-0.60(1)\pi$ corresponds to the global twist phase of $\theta=0$ (the determination of which peak corresponds to $0$ and which to $\pi$ is made by the value of $P_6$ at a slightly later time, $\sim$1.2~$\mu$s).
For the experimental implementation, we directly account for a calibration parameter $\theta_c=-0.60\pi$ in the IF output program to achieve a desired $\theta$ value at the atoms. 

\subsection{Additional state dynamics}

In the context of single-atom AB caging, we primarily present data for sites $\ket{5}$ and $\ket{6}$ in the main text. However, we measured population dynamics for all the states in the vicinity of the initially populated $\ket{6}$ - namely, states $\ket{4}$, $\ket{5}$, $\ket{6}$, $\ket{7}$, and $\ket{8}$. 
Figure~\ref{FIG:figs2} presents the extended dynamics data set for all of these state populations, for the same phase combinations of $\phi$ and $\theta$ as presented in Fig.~2 of the main text. Overall, all the state population dynamics are in good agreement with the theory predictions.
For the cases of AB caging [$\phi = \pi$, panels (b,d)], where initialization at site $\ket{6}$ should lead simply to a breathing dynamics between occupation of the $\ket{6}$ site and equal-weight occupation at the sites $\ket{4}$, $\ket{5}$, $\ket{7}$, and $\ket{8}$, we indeed observe nearly identical dynamics at each of these outer sites of the CLS.

\section{ADDITIONAL THEORETICAL DETAILS}

\subsection{Modeling the dipolar interactions}

In the main text we work with two different level structures: a 12-level twisted rhombic lattice (see Fig. 1) and a 6-level bipyramid (see Fig. 4). 
As discussed in~\cite{chen2024}, the full interaction Hamiltonian for a pair of atoms ($A$ and $B$) is
\begin{equation}
 H_{\rm int} = \sum_{i,j}\sum_{i',j'}e^{i\Delta_{ij}^{i'j'}t}V_{iji'j'}\ket{i}_{\rm A}\bra{j}\otimes\ket{j'}_{\rm B}\bra{i'} + {\rm h.c.} \ ,
\end{equation}
with $V_{iji'j'}$ the dipolar interaction strength (labelled as $V_{ij}$ when $i=i',j=j'$ below) and $\hbar\Delta^{i'j'}_{ij}$ the energy gap between state pairs $\ket{i}_{\rm A}\ket{j'}_{\rm B}$ and $\ket{i'}_{\rm B}\ket{j}_{\rm A}$.

In the weak interaction regime with large interatomic distance, i.e., $V_{iji'j'} \ll |\Delta^{i'j'}_{ij}|$, the non-resonant state-changing interaction terms only slightly affect the dynamics. We can thus largely neglect such terms and only consider the resonant state-conserving terms in ``ideal'' numerical simulations. 
Table~\ref{tabs1} lists all the relevant $C_3$ coefficients for resonant dipolar exchange interactions in both systems (the 12-state and 6-state lattices). We have previously calibrated the interaction strength for state pair $\ket{6}\leftrightarrow\ket{7}$ as $V_{67}=h\times 0.86(2)~{\rm MHz}$, based on a $2~{\rm MHz}$ separation between the frequency tones driving the acousto-optic deflector (AOD) used to create our optical tweezer pattern (corresponding to $\sim 9.6~\mu{\rm m}$ for the interatomic distance)~\cite{chen2024}. In our experiment, to obtain different $V/\Omega$ ratios, we vary the interaction strength by directly changing the frequency gap of the tones driving the AOD or we vary the microwave-driven hopping rates by globally changing the microwave coupling strength. For numerical simulations, we scale the interaction strengths $V_{ij} \propto C_3^{ij} / 2R_{\rm AB}^3$ with the calculated $C_3^{ij}$ coefficients and the interatomic distance $R_{\rm AB}$ based on the calibrated $V_{67}$ value. 

\begin{table*}[b]
\begin{tabular}{l|c||l|c}
 \hline
 \multicolumn{4}{l}{\textbf{Twisted rhombic lattice}}\\
 \hline
 $\ket{i}\ket{j}$ & $C_3$ & $\ket{i}\ket{j}$ & $C_3$ \\\hline
 $\ket{0}\ket{1}$ & -355.6 & $\ket{6}\ket{4}$ & -375.6 \\
 $\ket{0}\ket{2}$ & -705.0 & $\ket{6}\ket{5}$ & -756.4 \\
 $\ket{0}\ket{10}$ & 1422.3 & $\ket{6}\ket{7}$ & 1502.4 \\
 $\ket{0}\ket{11}$ & 705.0 & $\ket{6}\ket{8}$ & 756.4 \\\hline
 $\ket{3}\ket{1}$ & -414.3 & $\ket{9}\ket{4}$ & 1289.5 \\
 $\ket{3}\ket{2}$ & -834.4 & $\ket{9}\ket{5}$ & 639.1 \\
 $\ket{3}\ket{10}$ & 1567.3 & $\ket{9}\ket{7}$ & -322.4 \\
 $\ket{3}\ket{11}$ & 834.4 & $\ket{9}\ket{8}$ & -639.1 \\\hline
 $\ket{3}\ket{4}$ & -322.4 & $\ket{9}\ket{1}$ & 1657.3 \\
 $\ket{3}\ket{5}$ & -639.1 & $\ket{9}\ket{2}$ & 834.4\\
 $\ket{3}\ket{7}$ & 1289.5 & $\ket{9}\ket{10}$ & -414.3 \\
 $\ket{3}\ket{8}$ & 639.1 & $\ket{9}\ket{11}$ & -834.4 \\\hline
 \hline
 \multicolumn{4}{l}{\textbf{Bipyramid structure}}\\
 \hline
 $\ket{i}\ket{j}$ & $C_3$ & $\ket{i}\ket{j}$ & $C_3$ \\\hline
 $\ket{1}\ket{2}$ & -375.6 & $\ket{6}\ket{2}$ & 1502.4 \\
 $\ket{1}\ket{3}$ & 1502.4 & $\ket{6}\ket{3}$ & -375.6 \\
 $\ket{1}\ket{4}$ & -756.4 & $\ket{6}\ket{4}$ & 756.4 \\
 $\ket{1}\ket{5}$ & 756.4 & $\ket{6}\ket{5}$ & -756.4 \\\hline
\end{tabular}
\caption{\textbf{Calculated $C_3$ coefficients (units of $\rm{MHz} \ \mu{\rm m}^3$) for the resonant dipolar exchange interaction terms.} The table lists the relevant interaction terms for both structures (12-state lattice and 6-state bipyramind) described in the main text.
\label{tabs1}
}
\end{table*}

For strong interactions with relatively small interatomic distances (i.e., for rather large $V/\Omega$ ratios), the state-changing terms largely account for the differences between the ideal numerical simulations and experimental measurements in the main text Fig.~3(e).
To note, since the ratio $V/\Omega \lesssim 1$ for all the data relating to the 3D bipyramid structure (main text Fig.~4), and the state-changing transitions do not play a significant role in that case, here we focus our discussion only on the the case of the 12-level rhombic lattice structure.
To address the effects from non-resonant state-changing interaction terms, we include all the Zeeman sublevels with $\ket{0}$ to $\ket{11}$ as defined in Fig.~1, along with an additional 6 states that are in fact not part of our synthetic lattice, which we define as: $\ket{12}=\ket{44S_{1/2}, m_J=-1/2}$, $\ket{13}=\ket{43P_{3/2},m_J=-3/2}$, $\ket{14}=\ket{43P_{3/2},m_J=3/2}$, $\ket{15}=\ket{42P_{3/2},m_J=-3/2}$, $\ket{16}=\ket{42P_{3/2},m_J=3/2}$, $\ket{17}=\ket{42S_{1/2},m_J=-1/2}$. With a quantization $B$-field of $27~G$ (along $\hat{z}$, with the interatomic axis along $\hat{x}$), the $C_3$ coefficients and $\Delta_{ij}^{i'j'}$ for all additional resonant state-conserving and non-resonant state-changing interaction terms are listed in Table~\ref{tabs2}. In these full numerical simulations, we again scale all interaction strengths $V_{iji'j'}\propto C_3^{iji'j'}/2R^3_{\rm AB}$ to the calibrated value for $V_{67}$.

\subsection{Accounting for parameter uncertainties and the thermal spread of the interatomic separations}

In our full simulations (dashed lines), we also include confidence intervals (shaded regions) that reflect our uncertainties of the calibrated system parameters as well as the shot-to-shot position variance of the finite-temperature atoms due to trap release; see details of the Monte Carlo simulations used to model our confidence intervals in Ref.~\cite{chen2024}. We note that the shaded confidence regions shown in Figs.~3(d,e) and Fig.~4(f) are largely determined by the spread/uncertainty of the interaction strengths that results from the thermal spread of interatomic distances.

\begin{table*}[t]
\begin{tabular}{l|c|c||l|c|c||l|c|c}
 \hline
 \multicolumn{9}{l}{\textbf{Non-resonant state changing terms}}\\
 \hline
 $\ket{i}\ket{j}\ket{i'}\ket{j'}$ & $\Delta^{i'j'}_{ij}$ & $C_3$ & $\ket{i}\ket{j}\ket{i'}\ket{j'}$ & $\Delta^{i'j'}_{ij}$ & $C_3$ & $\ket{i}\ket{j}\ket{i'}\ket{j'}$ & $\Delta^{i'j'}_{ij}$ & $C_3$\\\hline
 $\ket{6}\ket{4}\ket{15}\ket{17}$ & 25 & -650.5 & $\ket{6}\ket{7}\ket{4}\ket{17}$ & 25 & 1502.4 & $\ket{6}\ket{16}\ket{7}\ket{17}$ & 25 & -650.5 \\
 $\ket{6}\ket{4}\ket{7}\ket{17}$ & 125 & 1126.8 & $\ket{6}\ket{16}\ket{15}\ket{17}$ & -75 & 3380.3 & $\ket{6}\ket{16}\ket{4}\ket{6}$ & -100 & 1951.6 \\
 $\ket{17}\ket{15}\ket{7}\ket{17}$ & -100 & 1951.6 & $\ket{6}\ket{8}\ket{5}\ket{17}$ & 50 & -756.4 & $\ket{17}\ket{8}\ket{5}\ket{6}$ & -100 & -2269.1 \\\hline
 $\ket{3}\ket{4}\ket{15}\ket{9}$ & 25 & -558.4 & $\ket{3}\ket{7}\ket{4}\ket{9}$ & 25 & 1289.5 & $\ket{3}\ket{16}\ket{7}\ket{9}$ & 25 & -558.4 \\
 $\ket{3}\ket{4}\ket{7}\ket{9}$ & 125 & 967.1 & $\ket{3}\ket{16}\ket{15}\ket{9}$ & -75 & 2901.4 & $\ket{3}\ket{16}\ket{4}\ket{3}$ & -100 & 1675.1 \\
 $\ket{9}\ket{15}\ket{7}\ket{9}$ & -100 & 1675.1 & $\ket{3}\ket{8}\ket{5}\ket{9}$ & 50 & -639.1 & $\ket{9}\ket{8}\ket{5}\ket{3}$ & -100 & -1917.4 \\\hline
 $\ket{3}\ket{1}\ket{13}\ket{9}$ & 25 & -717.6 & $\ket{3}\ket{10}\ket{1}\ket{9}$ & 25 & 1657.3 & $\ket{3}\ket{14}\ket{10}\ket{9}$ & 25 & -717.6 \\
 $\ket{3}\ket{1}\ket{10}\ket{9}$ & 125 & 1243.0 & $\ket{3}\ket{14}\ket{13}\ket{9}$ & -75 & 3729.0 & $\ket{3}\ket{14}\ket{1}\ket{3}$ & -100 & 2452.9 \\
 $\ket{9}\ket{13}\ket{10}\ket{9}$ & -100 & 2452.9 & $\ket{3}\ket{11}\ket{2}\ket{9}$ & 50 & -834.4 & $\ket{9}\ket{11}\ket{2}\ket{3}$ & -100 & -2503.2 \\\hline
 $\ket{0}\ket{1}\ket{13}\ket{12}$ & 25 & -615.9 & $\ket{0}\ket{10}\ket{1}\ket{12}$ & 25 & 1422.3 & $\ket{0}\ket{14}\ket{10}\ket{12}$ & 25 & -615.9 \\
 $\ket{0}\ket{1}\ket{10}\ket{12}$ & 125 & 1066.7 & $\ket{0}\ket{14}\ket{13}\ket{12}$ & -75 & 3200.2 & $\ket{0}\ket{14}\ket{1}\ket{0}$ & -100 & 1847.7 \\
 $\ket{12}\ket{13}\ket{10}\ket{12}$ & -100 & 1847.7 & $\ket{0}\ket{11}\ket{2}\ket{12}$ & 50 & -705.0 & $\ket{12}\ket{11}\ket{2}\ket{0}$ & -100 & -2114.9 \\\hline
 \hline
 \multicolumn{9}{l}{\textbf{Additional resonant terms}}\\
 \hline
 $\ket{i}\ket{j}$ & ~ & $C_3$& $\ket{i}\ket{j}$ & ~ & $C_3$ \\\hline
 $\ket{0}\ket{14}$& ~ & -1066.7 & $\ket{17}\ket{7}$& ~ & -375.6 \\
 $\ket{3}\ket{14}$& ~ & -1243.0 & $\ket{17}\ket{8}$& ~ & -756.4 \\
 $\ket{9}\ket{13}$& ~ & -1243.0 & $\ket{17}\ket{15}$& ~ & -1126.8 \\
 $\ket{3}\ket{16}$& ~ & -967.1 & $\ket{12}\ket{1}$& ~ & 1422.3 \\
 $\ket{9}\ket{15}$& ~ & -967.1 & $\ket{12}\ket{2}$& ~ & 705.0 \\
 $\ket{6}\ket{16}$& ~ & -1126.8 & $\ket{12}\ket{10}$& ~ & -355.6 \\
 $\ket{17}\ket{4}$& ~ & 1502.4 & $\ket{12}\ket{11}$& ~ & -705.0 \\
 $\ket{17}\ket{5}$& ~ & 756.4 & $\ket{12}\ket{13}$& ~ & -1066.7 \\\hline
\end{tabular}
\caption{\textbf{Calculated $C_3$ coefficients (units of $\rm{MHz} \ \mu{\rm m}^3$) and state pair energy differences $\Delta_{ij}^{i'j'}$ (units of $2\pi\times {\rm MHz}$) for all relevant dipolar exchange interaction terms by taking all Zeeman sublevels into consideration for the twisted rhombic lattice.}
\label{tabs2}
}
\end{table*}

\subsection{AB caging in the rhombic bipyramid}

Aside from viewing it as a two-unit cell (with periodic boundaries) realization of the diamond lattice, the energy structure and AB caging of the rhombic bipyramid lattice (Fig.~4 in the main text) can be analyzed by viewing it as a bi-partite lattice with a chiral symmetry. Sublattice $A$ is made of sites $\ket{1}$ and $\ket{6}$, and sublattice $B$ is made of sites $\ket{2}$, $\ket{3}$, $\ket{4}$, and $\ket{5}$. Since sublattice $B$ has two states more than sublattice $A$, there must be two zero modes. Since there is a chiral symmetry, the entire spectrum will be particle-hole symmetric; if there is a state with energy $E$, one obtains another state with energy $-E$ by adding minus signs ($\pi$ phase shifts) to the wavefunction amplitudes on the $A$ sublattice sites. Now what is left is to understand the two-fold degeneracy at nonzero energies. This part is where the Aharonov--Bohm (AB) caging becomes relevant. First of all, nonzero energy states should have some wavefunction on both the $A$ and $B$ sublattices. Because of the $\pi$-flux through the faces of the bipyramid, there is a state which has nonzero wavefunction amplitude on site $\ket{1}$ but zero wavefunction on site $\ket{6}$. Because of the up-down symmetry of the structure (up to a gauge transformation that changes the location of the phase-inverted hopping terms), there is a state with the same energy whose wavefunction on site $\ket{6}$ is nonzero but is zero on site $\ket{1}$. (These two states are linearly independent because one has no wave function on site $\ket{6}$ but the other has nonzero wavefunction on site $\ket{6}$.) Thus the degeneracy at nonzero energy is the result of the Aharonov-Bohm caging. To note, the degeneracy at zero energy is unrelated to Aharonov-Bohm caging, appearing for all values of the flux on the faces of the bipyramid structure.

\subsection{Derivation of emergent magnetism from interacting flat bands}

The non-interacting single particle Hamiltonian for the rhombic bipyramid structure shown in Fig.~4(b) of the main text reads
\begin{equation}\label{eqS2}
 H_{\rm sp} = \frac{\Omega}{2}\sum\nolimits_{j=2}^{5} (\ket{1} \bra{j} + e^{i\phi_{6j}}\ket{j}\bra{6}) + {\rm h.c.}
\end{equation}
with $\phi_{62}=\phi_{64} = \pi$ and $\phi_{63}=\phi_{65}=0$. As shown in Fig.~4(c) of the main text, this Hamiltonian has three sets of degenerate pairs of CLSs: 
(i) The upper $+\Omega$ energy band with wavefunctions $\ket{\lambda_u^-} = \frac{1}{\sqrt{8}}(-2|1\rangle - |2\rangle - |3\rangle - |4\rangle - |5\rangle)$ and $\ket{\lambda_u^+} = \frac{1}{\sqrt{8}}(-2|6\rangle + |2\rangle - |3\rangle + |4\rangle - |5\rangle)$.
(ii) The middle zero energy band with $\ket{\lambda_0^-} = \frac{1}{10}( - |2\rangle - 7|3\rangle + |4\rangle +7 |5\rangle)$ and $\ket{\lambda_0^+} = \frac{1}{10}( -7|2\rangle + |3\rangle + 7|4\rangle - |5\rangle)$.
(iii) The lower $-\Omega$ energy band with $\ket{\lambda_l^-} = \frac{1}{\sqrt{8}}(2|1\rangle - |2\rangle - |3\rangle - |4\rangle - |5\rangle)$ and $\ket{\lambda_l^+} = \frac{1}{\sqrt{8}}(2|6\rangle + |2\rangle - |3\rangle + |4\rangle - |5\rangle)$.
Relevant for the description of emergent ground state magnetism, one can encode an effective spin-1/2 into the two degenerate states in the lowest flat band as [see Figs.~4(c,d)]
\begin{eqnarray}\label{eqS3}
 \ket{\downarrow} &=& \ket{\lambda_l^-},\nonumber\\
 \ket{\uparrow} &=& \ket{\lambda_l^+}.
\end{eqnarray}
To note, the states $\ket{\downarrow}$ and $\ket{\uparrow}$ are also referred to in the main text as $\ket{\Psi_\downarrow}$ and $\ket{\Psi_\uparrow}$, respectively.

To obtain the effective magnetic Hamiltonian for an atom pair (labelled as $A$ and $B$), we expand the interaction Hamiltonian
\begin{equation}
 H_{\rm int} = \sum\nolimits_{j=2}^{5} \left(V_{6j} \ket{6}_{\rm A}\bra{j} \otimes \ket{j}_{\rm B}\bra{6} + V_{1j} \ket{1}_{\rm A}\bra{j} \otimes \ket{j}_{\rm B}\bra{1}\right) + {\rm h.c.}
\end{equation}
under the spin basis $\{\ket{\downarrow\downarrow}, \ket{\downarrow\uparrow}, \ket{\uparrow\downarrow}, \ket{\uparrow\uparrow}\}$. For uniform interaction strength, i.e., $V_{6j}=V_{1j}=V$ for $j\in [2,5]$, in the weak interaction limit, it takes the following form
\begin{equation}\label{eqS4}
 H_{\rm int} = \frac{V}{4}(\sigma_{\rm A}^z\sigma_{\rm B}^z + \mathbb{1}_{\rm A}\mathbb{1}_{\rm B}) + \frac{V}{4}(\sigma_{\rm A}^x\sigma_{\rm B}^x + \sigma_{\rm A}^y\sigma_{\rm B}^y).
\end{equation}
In this ideal case, the effective spin exchange rate is $2 J_{\rm eff,id} = 2J_{\rm XX}$, with $J_{\rm eff,id} = J_{\rm XX} = J_{\rm ZZ} = V/4$ being the common prefactor of the XX, YY, and ZZ terms. To note, the constant $\mathbb{1}_{\rm A}\mathbb{1}_{\rm B}$ term does not affect the physics of this system, and is omitted in the main-text description of the effective spin model. With the initial population all in state $\ket{6}$ (the initial condition in our experiment), the dynamics of this ideal model follows
\begin{equation}\label{eqS5}
 P_6(t) = \frac{1}{2} + \frac{1}{2}\cos{\left(\frac{2J_{\rm eff, id}}{\hbar}t\right)}\cos{\left(\frac{2\Omega}{\hbar}t\right)},
\end{equation}
which is actually a beating formula between two oscillation frequencies: 
\begin{enumerate}
    \item[(i)] one is the inter-band oscillation from the single particle Hamiltonian $H_{\rm sp}$, since the initial state $|6\rangle$ is a superposition of $\ket{\lambda_l^+}$ and $\ket{\lambda_u^+}$, i.e., $\ket{6} = \frac{1}{\sqrt{2}}(\ket{\lambda_l^+}-\ket{\lambda_u^+})$. These two eigenstates have an energy difference of $2\Omega$, which naturally makes the population in $\ket{6}$ oscillate with a frequency of $2\Omega/\hbar$.
    \item[(ii)] another is the interaction-induced effective spin exchange with the rate $2J_{\rm eff,id}/\hbar$.
\end{enumerate}

For our real, physical system shown in Fig.~4(a) of the main text, the interaction strengths of the various Rydberg state pairs are not uniform (in the synthetic dimension). We scale all of them to the strongest one, i.e., letting $V_{62}=V$, leading to $\{V_{62}, V_{63}, V_{64}, V_{65}\} = \{V, -V/4, V/2, -V/2\}$ and $\{V_{12}, V_{13}, V_{14}, V_{15}\} = \{-V/4, V, -V/2, V/2\}$ from the calculated $C_3$ coefficients. This experimental interaction Hamiltonian can be expanded in spin basis as
\begin{equation}\label{eqS6}
    H_{\rm int} = \frac{3V}{64}(\sigma_{\rm A}^z\sigma_{\rm B}^z + \mathbb{1}_{\rm A}\mathbb{1}_{\rm B}) + \frac{3V}{64}(\sigma_{\rm A}^x\sigma_{\rm B}^x + \sigma_{\rm A}^y\sigma_{\rm B}^y) + \frac{9V}{64}(\sigma_{\rm A}^x\sigma_{\rm B}^z + \sigma_{\rm A}^z\sigma_{\rm B}^x).
\end{equation}
The effective spin exchange rate is $2J_{\rm XX}$ with $J_{\rm XX} = J_{\rm ZZ} = 3V/64$. However, the interaction Hamiltonian also includes an additional XZ term, which induces correlated spin-flip oscillations
($\ket{\uparrow\uparrow}\leftrightarrow -\ket{\uparrow\downarrow}$, $\ket{\uparrow\uparrow}\leftrightarrow -\ket{\downarrow\uparrow}$,
$\ket{\downarrow\downarrow}\leftrightarrow\ket{\uparrow\downarrow}$, and
$\ket{\downarrow\downarrow}\leftrightarrow\ket{\downarrow\uparrow}$)
that consequently affect the $P_6$ dynamics.
By comparing Hamiltonians (\ref{eqS6}) and (\ref{eqS4}), the beating dynamics should be modified [with the appropriate form given in the caption of Fig.~\pref{FIG:figs3}{d}], with multiple low-frequency contributions owing to the added presence of the $J_{\rm XZ}$ terms.
In our experiment, we simply fit the $P_6$ dynamics to the form $P_6(t) = a + b \cos{\left(\frac{2\omega}{\hbar}t\right)}\cos{\left(\frac{2\Omega}{\hbar}t\right)}$ inspired by Eq.~(\ref{eqS5}).
These fits yield a trend of $\omega$ values that is in fair agreement with the analytical expression for the effective interaction scale that accounts for the influence of the $J_{\rm XZ}$ terms, given by $J_{\rm eff} = \sqrt{J^2_{\rm XX} + J^2_{\rm XZ}/2} \propto V$. This approximate form is shown as the solid line in Fig.~4(g) of the main text.
We find that the experimentally observed rates $\omega$ are also in good agreement with the expected response based on analogous fits of the numerically simulated dynamics, which is shown as the dashed-dotted line in Fig.~4(g). To note, the linear scaling of $J_{\rm eff}$ with $V$ is expected to break down due to band mixing as $V$ approaches $\Omega$, which is reflected in the prediction based on numerical simulations.

\begin{figure*}[]
	\includegraphics[width=\textwidth]{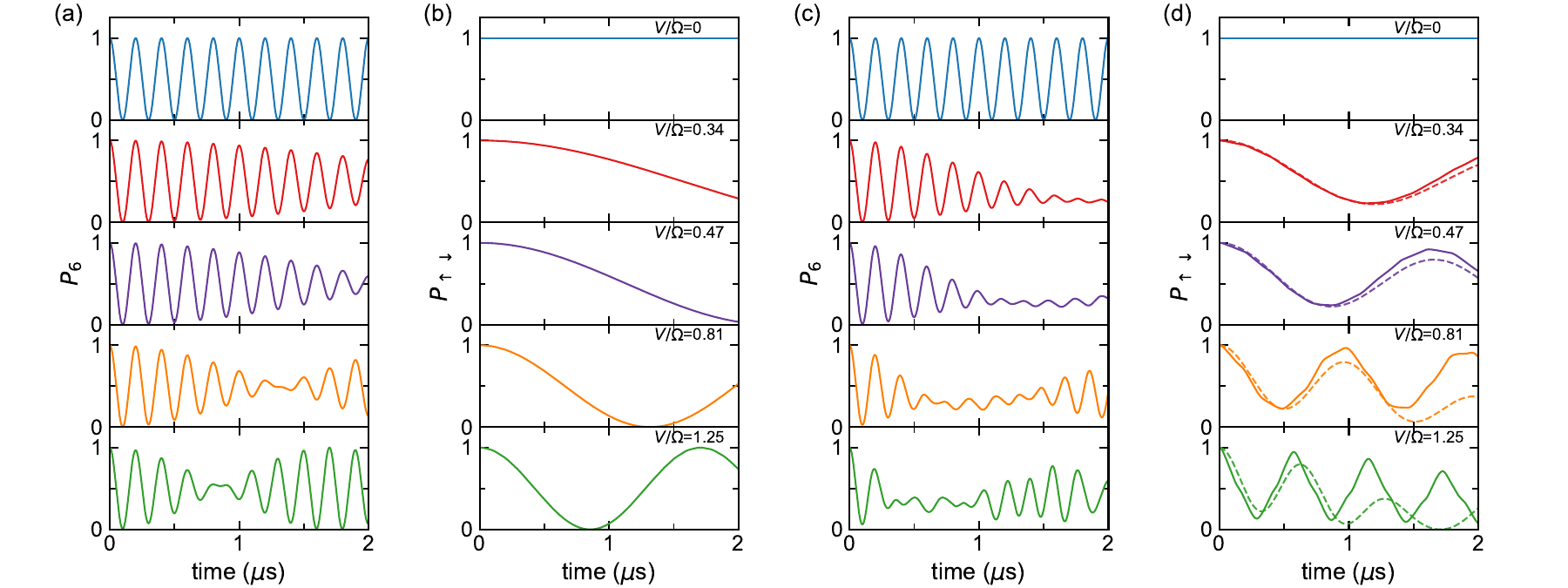}
        \centering
	\caption{\label{FIG:figs3}
        \textbf{Comparison of the dynamics based on Hamiltonians (\ref{eqS2}) and (\ref{eqS4}): (a,b) with a uniform interaction $3V/16$ which leads to Hamiltonian (\ref{eqS6}) excluding the $\sigma_x\sigma_z$ term; (c,d) with the real interaction used in the experiment, i.e., Eq.~(\ref{eqS6}) with all interaction terms.}
        \textbf{(a)}~``Ideal model'' time evolution of the population in $\ket{6}$ under different interactions, with the initial state $\ket{6}_{\rm A}\ket{6}_{\rm B}$.
        \textbf{(b)}~``Ideal model'' time evolution of the population in $\ket{\uparrow\downarrow}$, with the initial state $\ket{\uparrow\downarrow}$=$\ket{\lambda_l^+}_{\rm A}\ket{\lambda_l^-}_{\rm B}$; see (\ref{eqS3}). With no fast oscillations (because the population is largely restricted to a single band), the dynamics is fully determined by
        $P_{\uparrow\downarrow}(t)=\cos^2{(2J_{\rm XX}t/\hbar)}=\cos^2{(2J_{\rm eff, id}t/\hbar)}$
        with the effective spin-exchange interaction strength $J_{\rm eff, id} = J_{\rm XX}=3V/64$. 
        \textbf{(c)}~``Physical model'' time evolution of the population in $\ket{6}$ under different interactions, with the initial state $\ket{6}_{\rm A}\ket{6}_{\rm B}$. With additional effective $\sigma_x\sigma_z$ terms, the full dynamics are modified from the simple beating formula of (\ref{eqS5}). The $\sigma_x\sigma_z$ terms only change the envelope of the beating; the fast dynamics still have an oscillation frequency of $2\Omega/\hbar$, as seen by comparing (a) and (c). 
        \textbf{(d)}~``Physical model'' time evolution, with initial conditions and plotted observable similar to those in panel (b). The added $\sigma_x\sigma_z$ terms increase the overall rate of intra-band pseudospin dynamics. By diagonalizing the effective spin Hamiltonian (\ref{eqS6}), we find that the dynamics should follow the formula $P_{\uparrow\downarrow}(t) = \frac{3}{8}+\frac{1}{4}\cos{(2(J_{\rm XZ}-2J_{\rm XX})t/\hbar)} + \frac{1}{4}\cos{(2(J_{\rm XZ}+2J_{\rm XX})t/\hbar)}+\frac{1}{8}\cos{(4J_{\rm XZ}t/\hbar)}$
        [$\sim \cos^2{(2J_{\rm eff}t/\hbar)}$ in short time limit,
        with $J_{\rm eff}=\sqrt{J_{\rm XX}^2 + J_{\rm XZ}^2/2}$],
        shown as dashed lines. Short-time fits to Eq.~(\ref{eqS5}) should reveal a low frequency component $\omega \approx J_{\rm eff}$.
        For large $V$, the analytical results (dashed lines) deviate from the numerical simulation with the original Hamiltonian (solid lines), due to the onset of band mixing as $V$ approaches the band gap, $\Omega$ (see Fig.~4(e) of the main text).
        For all simulations here, we use $\Omega/h=2.5~{\rm MHz}$ and $V/\Omega=\{0, 0.34, 0.47, 0.81, 1.25\}$ from top to bottom panels. 
}
\end{figure*}

\clearpage
\end{widetext}
\end{document}